\documentclass{article}
\usepackage{LaThuileFPSpro}
\usepackage{graphicx}
\usepackage{epsfig}
\newcommand{\bo}{B^0}

\newcommand{\bs}{B_s}

\newcommand{\ks}{K_S}

\def\bbarb{B^0 -\bar B^0}
\def\OL{\overline}
\def\a{{\cal A}}

\def\A{{\cal A}}

\def\pr{{Phys. Rev.}~}
\def\prl{{ Phys. Rev. Lett.}~}
\def\pl{{ Phys. Lett. }~}
\def\np{{ Nucl. Phys.}~}

\def\lsim{ {\
\lower-1.2pt\vbox{\hbox{\rlap{$<$}\lower5pt\vbox{\hbox{$\sim$}
}}}\ } }
\def\gsim{ {\
\lower-1.2pt\vbox{\hbox{\rlap{$>$}\lower5pt\vbox{\hbox{$\sim$}
}}}\ } }

\def\be{\begin{eqnarray}}
\def\en{\end{eqnarray}}

\def\ov{\overline}
\def\hep{hep-ph}
\def\Ref{\bibitem}
\def\jpsi{J/\psi~}
\begin{document}
\def\op{{\cal O}}

\title{\bf CP VIOLATION HIGHLIGHTS: CIRCA 2005}
\author{Amarjit Soni\\
{\em High Energy Theory, Department of Physics,}\\
{\em Brookhaven National Laboratory, Upton, NY  11973-5000}\\
}

\maketitle

\baselineskip=11.6pt

\begin{abstract}
Recent highlights in CP violation phenomena are reviewed. B-factory
results imply that CP-violation phase in the CKM matrix is the
dominant contributor to the observed CP violation in K and B-physics.
Deviations from the predictions of the CKM-paradigm due to beyond the
Standard Model CP-odd phase are likely to be a small perturbation.
Therefore, large data sample of clean B's will be needed. Precise
determination of the unitarity triangle, along with time dependent
CP in penguin dominated hadronic and radiative modes are
discussed. {\it Null tests} in B, K and top-physics 
and separate determination of the K-unitarity triangle
are also  emphasized. 
\end{abstract}
\newpage
\section{B-factories help attain an important milestone: Good and bad news}
The two asymmetric B-factories at SLAC and KEK have provided a
striking confirmation of the CKM paradigm~\cite{ckm}. Existing experimental
information from the indirect CP violation
parameter, $\epsilon$ for the
$K_{L}\rightarrow \pi \pi$, semileptonic
$b\rightarrow ue\nu$ and $\bbarb$ mixing along with lattice calculations
predict that in the SM, $(\sin2\beta)\simeq.70\pm.10$~\cite{ans01,ckmfit,utfit}.
This is in very
good agreement with the BELLE and BABAR result~\cite{hfag04}:\\
\be
\a_{CP}(B^0\to \psi K^0) = \sin2\beta =.726\pm.037  
\label{psik}
\en
This leads to the conclusion that the CKM phase of the 
Standard Model (SM) is the
dominant contributor to $\a_{CP}$. That, of course, also means that
CP-odd phase(s) due to beyond the Standard Model (BSM) sources may
well cause only small deviations from the SM in B-physics.  \\
Actually, there are several reasons to think that BSM phase(s) may
cause only small deviations in B-physics. In this regard, SM itself
teaches a very important lesson.

\section{Important lesson from the CKM-paradigm}
We know now that the CKM phase is 0(1) (actually, the 
CP violation parameter $\eta$ is
0(.3)~\cite{ans01,ckmfit,utfit}). 
The CP effects that it causes on different observables though
is quite different. In K-decays, the CP asymmetrics are $\le10^{-3}$. In
charm physics, also there are good reasons to expect small observable
effects. In top physics, the CKM phase causes completely negligible
effects~\cite{ehs91,abes_pr}. Thus only in B-decays, the large asymmetries
(often 0(1)) are caused by the CKM phase. So  even if the BSM
phase(s) are 0(1) it is unlikely that again in B-physics they will
cause large effects just as the SM does.


%
\section{Remember the $m_{\nu}$}
Situation with regard to BSM CP-odd phase(s) ($\chi_{BSM}$) is somewhat
reminiscent of the neutrino mass ($m_{\nu}$)~\cite{brwns}. 
There was no good
reason for $m_{\nu}$ to be zero; similarly, there are none for
$\chi_{BSM}$ to be zero either.
In the case of $\nu^{'}s$, there were the solar $\nu$ results that were
suggestive for a very long time; similarly, in the case of $\chi_{BSM}$,
the fact that in the SM, baryogenesis is difficult to accomodate
serves as the beacon.\\
It took decades to show $m_{\nu}$ is not zero: $\Delta m^{2}$ had to be
lowered from $\sim O(1-10)eV^{2}$ around 1983 down to
$O(10^{-4}eV^{2})$
  before $m_{\nu} \neq0$ was established via neutrino oscillations. We can
  hope for better luck with $\chi_{BSM}$ but there is no good reason to
  be too optimistic; therefore, we should not rely on luck but rather
  we should seriously prepare for this possibility.\\

%
To recapitulate, just as the SM-CKM phase is 0(1), but it caused only
$0(10^{-3})$ CP symmetries in K- decays, an 0(1) BSM-CP-odd phase may
well cause only very small asymmetries in B-physics. To search for
such small effects:\\
1) We need lots and lots of clean B's ({\it i.e.} $0(10^{10}$) or more)
\\
2) Intensive study of $B_{s}$ mesons (in addition to B's) becomes very
important as comparison between the two types of B-mesons will teach
us how to improve quantitative estimates of flavor symmetry breaking
effects.
\\
3) We also need clean predictions from theory (wherein item 2 should
help).
\section{Improved searches for BSM phase}
Improved searches for BSM-CP-odd phase(s) can be subdivided into the
following main categories:\\
\\
a) Indirect searches with theory input\\
b) Indirect searches without theory input\\
c) Direct searches.\\

\subsection{Indirect searches with theory input}

Among the four parameters of the CKM
matrix, $\lambda, A,\rho$ and $\eta$,
$\lambda = 0.2200 \pm 0.0026$, $A \approx 0.850 \pm 0.035$
\cite{PDB}
are known quite precisely; $\rho$ and $\eta$ still need to be
determined accurately. Efforts have been underway for many
years to determine these parameters. The angles $\alpha, \beta, \gamma$,
of the unitarity triangle (UT) 
can be determined once one knows the 4-CKM parameters.

A well studied strategy for determining these from experimental data
requires knowledge of hadronic matrix elements.
Efforts to calculate several
of the relevant matrix elements on the lattice, with increasing
accuracy, have been underway for past many years.
A central role is played by the following four
inputs~\cite{ans01,ckmfit,utfit}:

\begin{itemize}
\item $B_K$ from the lattice with $\epsilon$ from experiment
\item $f_B\sqrt{B_B}$ from the lattice with $\Delta m_d$ from
experiment
\item $ \xi $ from the lattice with $\frac{\Delta m_s}{\Delta m_d}$
from experiment
\item $\frac{b \to u l \nu}{b \to c l \nu}$
from experiment, along with input from
phenomenology especially  heavy quark symmetry as well
as the lattice.
\end{itemize}

As mentioned above, for the past few years, these inputs have led to the
important
constraint: $ \sin 2\beta_{SM} \approx 0.70 \pm 0.10$
which is found to be
in very good agreement with 
direct experimental determination, Eq.~\ref{psik}. 

Despite severe limitations (e.g. the so-called quenched approximation)
these lattice inputs provided 
valuable help so that with B-Factory measurements one arrives at the
very important
conclusion that in $B \to \jpsi K^0$ the CKM-phase is
the dominant contributor; any new physics (NP) contribution
is unlikely to be greater than about 15\%.

What sort of progress can we expect from the lattice in the next
several years in these (indirect) determination
of the UT? To answer this it is useful to look back and
compare where we were to where we are now. 
Perhaps this gives us an indication
of the pace of progress of the past several years.
Lattice calculations
of matrix elements around 1995~\cite{lat95_as}
yielded (amongst other things) $\sin 2\beta \approx 0.59 \pm
0.20$, whereas the corresponding error decreased to around
${\pm 0.10}$ around 2001~\cite{ans01}.
In addition to $\beta$,
such calculations also now constrain $\gamma (\approx 60^\circ)$
with an error of around $10^\circ$~\cite{ans01}.

There are three important developments that should help lattice
calculations in the near future:\\
\begin{enumerate}
\item{} Exact chiral symmetry can be maintained on the lattice.
This is especially important for light quark physics.
\item{} Relatively inexpensive methods for simulations
with dynamical quarks (esp. using improved staggered fermions~\cite{hpqcd})
have become
available. This should help overcome limitations of the
quenched approximation.
\item{} About  a factor of 20 increase in computing power
is now being used compared to 
a few years ago. 
\end{enumerate}

As a specific example one can see that the error on $B_K$
with the 1st use of dynamical domain wall fermions~\cite{rbc_nf2} now
seems to be reduced by about a factor of two~\cite{rbc_jun}. 
In the next few years or so
errors on lattice determination of CKM parameters should
decrease
appreciably, perhaps by a factor of 3.
So the error in $\sin 2\beta_{SM} \pm 0.10 \to
\pm 0.03$;
$\gamma \pm 10^\circ \to 4^\circ$ etc.
While this increase in accuracy is very welcome,
and will be very useful, 
there are good reasons to believe, experiment will
move ahead of theory in direct determinations
of unitarity angles in the next 5 years. (At present, experiment is already
ahead
of theory for $\sin 2 \beta$).

\subsection{Indirect searches without theory input: Elements of a
superclean UT}

One of the most exciting developments of recent years in B-physics
is that methods have been developed so that all three angles
of the UT can be determined cleanly with very small theory errors.
This is very important as it can open up several ways
to test the SM-CKM paradigm of CP violation; in particular, the
possibility
of searching for small deviations. Let us very briefly recapitulate
the methods in question:\\

\begin{itemize}

\item Time dependent CP asymmetry (TDCPA) measurements in $B^0, \bar B^0
\to \psi K^0$ type of final states should give the angle $\beta$ very
precisely with an estimated irreducible theory error (ITE)
of $\le O(0.1\%)$~\cite{bmr04}.

\item Direct CP (DIRCP) studies in $B^{\pm} \to ``K^{\pm}" D^0, \bar D^0$
gives $\gamma$ very\\
cleanly~\cite{ans_pristine,gsz05}.

\item TDCPA measurements in $B^0, \bar B^0 \to ``K^0" D^0, \bar D^0$
gives $(2 \beta + \gamma)$ {\it and} also $\beta$ very
cleanly~\cite{ans_b0,bgk05}.

\item In addition, TDCPA measurements in $B_s \to K D_s$
type modes also gives $\gamma$ very cleanly~\cite{pb_lhc}.

\item Determination of the rate for the CP violating decay
$K_L \to \pi^0 \nu \bar \nu$ is a very clean way to measure
the Wolfenstein parameter $\eta$, which is indeed the CP-odd
phase in the CKM matrix~\cite{ajb_nunu}.

It is important to note that the ITE for each of these methods is
expected to be $\le 1\%$, in fact perhaps even $\le 0.1\%$.

\item Finally let us briefly mention that, TDCPA studies of $B^0, \bar B^0
\to \pi \pi$ or $\rho \pi$ or $\rho \rho$ 
gives $\alpha$~\cite{glw,lnqs,alnq}. However, in
this case, isospin conservation needs be used and that requires,
{\it assuming} that electro-weak penguins (EWP) make negligible
contribution. This introduces some model dependence and
may cause an error of order a few degrees, {\it i.e.}
for $\alpha$ extraction the ITE may well end up being
O(a few \%). However,
given that there are three types of final states each of which
allows a determination of $\alpha$, it is quite likely
that further studies of these methods will lead to a
reduction of the common source of error
originating from isopsin violation due to the EWP.

It is extremely important that we make use of these opportunities
afforded to us by as many of these very clean redundant
measurements as possible. In order to exploit these methods to their
fullest  potential and get the angles with errors of order
ITE will, for sure, require a 
SUPER-B Factory(SBF)~\cite{brwns,sbf_kek,mn_04,sbf_slac}.

This in itself constitutes a strong enough reason for a SBF,
as it represents a great opportunity to precisely nail
down the important parameters of the CKM paradigm. 

\end{itemize}

\subsubsection{Prospects for precision determination of $\gamma$}

Below we briefly
discuss why the precision extraction of
$\gamma$ seems so promising.

For definiteness, let us recall the basic features of the
ADS method~\cite{ads1}. In this interference
is sought between two amplitudes of roughly similar size {\it i.e.}
$B^- \to K^- D^0$ and $B^- \to K^- \bar D^0$ where the $D^0$ and
$\bar D^0$ decay to common final states such as the simple
two body ones like $K^+ \pi^-$,
$K^+ \rho^-$, $K^+ a_1^-$, $K^{+*} \pi^-$ or they may also be multibody
modes e.g. the Dalitz decay $K^+ \pi^- \pi^0$, $K^+ \pi^- \pi^+ \pi^-$
etc.
It is easy to see that
the interference is between a colored allowed B decay followed by
doubly Cabibbo suppressed D decay and a color-suppressed B decay
followed by Cabibbo allowed D decay and consequently then interference
tends to be maximal and should lead to large asymmetries.

For a given (common) final state of $D^0$ and $\bar D^0$ the amplitude
involves three unknowns: the color
suppressed Br($B^- \to K^- \bar D^0$),
which is not directly accessible to experiment~\cite{ads1},
the strong phase $\xi_{f_i}^K$ and the weak phase $\gamma$.
Corresponding
to each such final state (FS) there are two observables: the rate for
$B^-$ decay and for the $B^+$ decay.

Thus, if you stick to just one common FS of $D^0$, $\bar D^0$,
you do not have enough information to solve for $\gamma$.
If you next consider two common FS of $D^0$ and $\bar D^0$
then you have one additional unknown (a strong phase) making
a total of 4-unknowns with also 4-observables. So with
two final states the system becomes soluble, i.e. we can then
use the experimental data to solve not only the value of
$\gamma$ but also the strong phases and the suppressed Br
for $B^- \to K^- \bar D^0$. With N common FS of $D^0$ and $\bar D^0$,
you will have 2N observables and N + 2 unknowns. We need
$2N \ge (N + 2) $ i.e. $N \ge 2$.
The crucial point, though, is that there are a very large number
of possible common modes of $D^0$ and $\bar D^0$ which can all be used
to improve the determination of $\gamma$. 

Let us briefly mention some of the relevant common
modes of $D^0$ and $\bar D^0$:

\begin{itemize}
\item The CP-eigenstate modes, originally discussed by
GLW~\cite{glw}: $K_S$ [$ \pi^0$, $ \eta$, $ \eta'$, $ \rho^0$, $ \omega$]; $ \pi^+ \pi^-$,....

\item CP-non-eigenstates (CPNES), discussed
by GLS~\cite{gls}: $K^{*+} K^-$, $ \rho^+ \pi^-$...
These are singly Cabibbo suppressed modes.

\item CPNES modes originally discussed by ADS~\cite{ads1,ads2}:
$K^{+(*)}$[$ \pi^-$, $\rho^-$, $a_1^-$....]

\item There are also many multibody modes, such as the Dalitz $D^0$ decays:
$K_S \pi^+ \pi^-$~\cite{ggsz} or $K^+ \pi^- \pi^0$~\cite{ads2} etc;
and also modes such as
$K^- \pi^+ \pi^- \pi^+$, $K^- \pi^+ \pi^- \pi^+ \pi^0$, or indeed
$K^- \pi^+ +n \pi$~\cite{ans_b0,ans_path,ans_charm}.
Furthermore, multibody modes such as $B^+ \to K_i^+ D^0 \to 
(K \pi)^+ D^0$ or $(K n \pi)^+ D^0$~\cite{ans_path,aegs}
can also be used.    

\end{itemize}

\begin{figure}[t]
\vspace{9.0cm}
\includegraphics{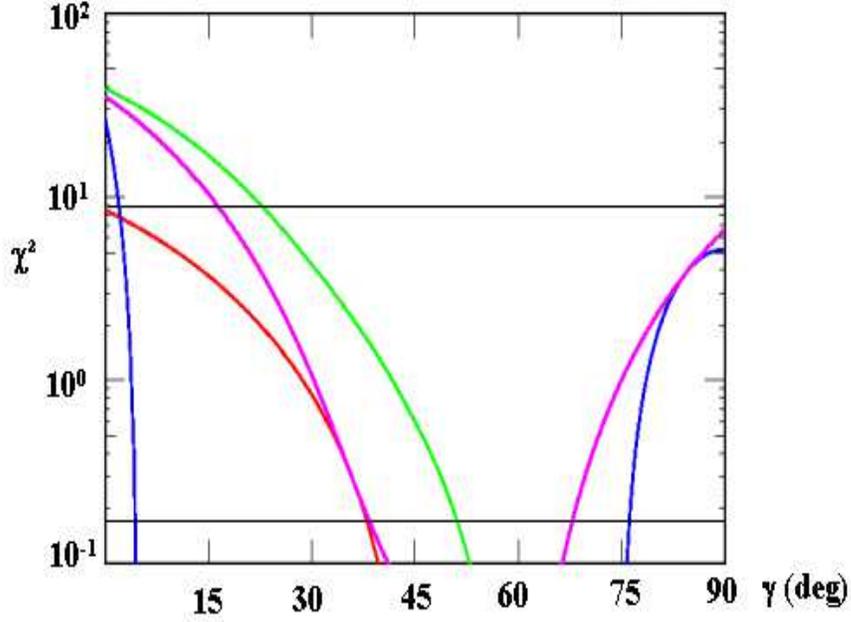}
\caption{\it
 $\gamma$ determination with incomplete input ({\it i.e.}
 cases when the number of observables is less than the number
 of unknown parameters). The upper horizontal line corresponds
 to low-luninosity {\it i.e.} around current B-factories
 whereas the lower horizontal curve is relevant for
 a SBF. Blue uses all CPES modes of $D^0$, 
red is with only $K^+ \pi^-$ and purple uses combination
 of the two. Green curve again uses on
 $D^0$, $\bar D^0 \to K^+ \pi^-$ but now includes $K^{*-}$ and 
 $D^{*0}$; see text for details.     
\label{un_det} }
\end{figure}

\begin{figure}[t]
\vspace{9.0cm}
\includegraphics{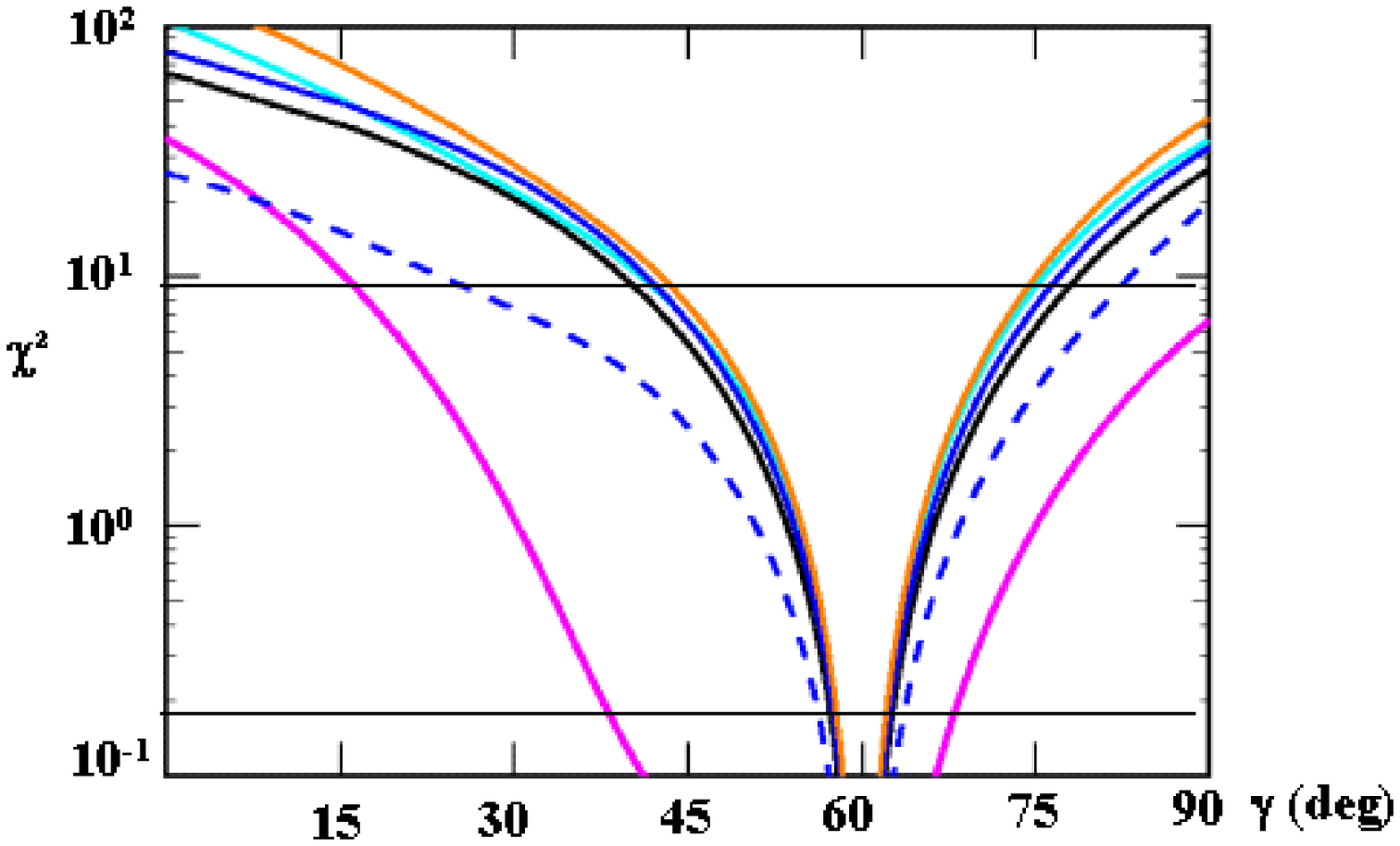}
\caption{\it
  $\gamma$ extraction with over-determined cases. Purple curve shows
 the effect
  of combining GLW (all CPES modes) with one ADS ($K^+ \pi^-$) mode; black
  curve differs from purple only in that it also
  includes $D^0$ from $D^{*0}$;
  blue curves show the effect of properly including the correlated
  strong phase between $D^{0*} \to D^0 + \pi$ and
  $D^{0*} \to D^0 + \gamma$. Orange curve includes lot
  more input including Dalitz and multibody modes.
  see text for details (See also Fig~\ref{un_det}). 
  Adopted from~\cite{ans_path}.   
\label{over_det} }
\end{figure}

Fig.~\ref{un_det} and Fig.~\ref{over_det} 
show how combining different strategies helps a great
deal. In the fig we show $\chi^2$ versus $\gamma$. As indicated above
when you consider an individual final state of $D^0$ and $\bar D^0$
then of course there are 3 unknowns ( the strong phase,
the weak phase ($\gamma$) and the ``unmeasureable" Br)
and only two observables (the rate for $B^-$ and the rate for $B^+$).
So in the figure, for a fixed value of $\gamma$,
we search for the minimum of the $\chi^2$ by letting
the strong phase and the ``unmeasureable" Br take any value they want.

Fig.~\ref{un_det} and Fig.~\ref{over_det}
show situtation with regard to under determined and
over determined cases respectively. The upper horizontal
line corresponds roughly to the low luminosity i.e. comparable
to the current B-factories\cite{sbf_kek,sbf_slac} whereas the lower horizontal
curve is relevant for a super B-factory. In Fig.~\ref{un_det} in blue
is shown the case when only the input from (GLW) CPES modes
of $D^0$ is used; note all the CPES modes are included here.
You see that the resolution on $\gamma$
then is very poor. In particular, this method is rather
ineffective in giving a lower bound; its upper bound
is better.




In contrast, a single ADS mode ($K^+ \pi^-$) is very effective
in so far as lower bound is concerned,
but it does not yield an effective upper bound (red).

Note that in these two cases one has only two observables and 3
unknowns.
In purple is shown the situation when these two methods are combined.
Then at least at high luminosity there is significant improvement
in attaining a tight upper bound; lower bound obtained by ADS alone
seems largely unaffected.

Shown in green is another under determined case consisting of
the use of a single ADS mode, though it includes $K^{*-}$ as well
$D^{*0}$; this again dramatically improves the lower bound.
From an examination of these curves it is easy to see that
combining information from different methods and modes
improves the determination significantly~\cite{ans_path}.

Next we briefly discuss some over determined cases (Fig.~\ref{over_det}).
In purple all the CPES modes of $D^0$ are combined with just
one doubly Cabibbo-suppressed (CPNES) mode. Here there are 4 observables
for the 4 unknowns and one gets a reasonable solution at least
especially for the high luminosity case.

The black curve is different from the purple one in only
one respect; the black one also includes the $D^{0*}$ from
$B^- \to K^- D^{0*}$ where subsequently the $D^{0*}$ gives
rise to a $D^0$. Comparison of the black one with the purple
shows considerable improvement by including the $D^{0*}$.
In this case the number of observables (8) exceeds
the number of unknowns (6).

Actually, the $D^{0*}$ can decay to $D^0$ via two modes:
$D^{0*} \to D^0 + \pi$ or $D^0 + \gamma$. Bondar and Gershon
\cite{bg}have
made a very nice observation that the strong phase for the
$\gamma$ emission is opposite to that of the $\pi$
emission. Inclusion (blue curves)
of both types of emission increases the number
of observables to 12 with no increase in number of unknowns.
So this improves the resolving power for $\gamma$ even more.

The orange curves show the outcome when a lot more input
is included; not only $K^-$, $K^{-*}$, $D^0$, $D^{0*}$
but also Dalitz and multibody decays of $D^0$ are included.
But the gains now are very modest; thus once the number of observables
exceeds the number of unknowns by a few (say O(3)) further
increase in input only has a minimal impact.

Let us briefly recall that  another important way to
get these angles is by studying time-dependent CP (TDCP)
(or mixing-induced CP (MIXCP)) 
violation via $B^0 \to D^{0(*)} ``K^0"$. Once again,
all the common decay modes of $D^0$ and $\bar D^0$
can be used just as in the case of direct CP studies involving $B^{\pm}$
decays.  Therefore, needless to say
input from charm factory~\cite{ans_charm,ggr,jr} 
also becomes desirable for MIXCP
studies of $B^0 \to D^{0(*)} ``K^0"$ as it is for direct CP
using $B^{\pm}$. It is important to stress that
this method gives not only the combinations of the angles
(2$\beta +\gamma \equiv \alpha - \beta + \pi$) but also
in addition this is another way to get $\beta$
cleanly~\cite{ans_b0,bgk05}.
In fact whether one uses $B^{\pm}$ with DIRCP or $B^0- \bar B^0$ with
TDCP these methods are very clean with (as indicated above)
the ITE of $\approx 0.1\%$. However, the TDCP studies
for getting $\gamma$ (with the use of $\beta$
as determined from $\psi K_s$ ) is less efficient
than with the use of DIRCP involving $B^{\pm}$. 
Once we go to luminosities $\ge 1 ab^{-1}$, though,  
the two methods for $\gamma$ should become
competitive. Note that this method for getting $\beta$ is
significantly less efficient than from
the $\psi K_s$ studies~\cite{ans_b0}.

\begin{table}[t]
\centering
\caption{\it Projections for direct determination of UT.
 }
\vskip 0.1 in
\begin{tabular}{|l|c|c|c|c|} \hline
 & Now(0.2/ab) & 2/ab & 10/ab & ITE \\
\hline
\hline
$\sin 2 \phi_1$ & 0.037 & 0.015 & 0.015(?) & 0.001 \\
\hline
$\alpha(\phi_2)$ & $13^\circ$ & $4^\circ$(?) & $2^\circ$(?) & ~$1^\circ$(?) \\
\hline
$\gamma(\phi_3)$ & $ \pm 20^\circ \pm 10^\circ \pm 10^\circ$ & $5^\circ$ to  $2^\circ$ & $ < 1^\circ(?)$ & $0.05^\circ$
  \\
\hline
\end{tabular}
\label{ut_pros}
\end{table}

Table~\ref{ut_pros} summarizes the current status and expectations
for the near future for the UT angles. With the current O(0.4/ab)
luminosity
between the two B-factories, $\gamma \approx (69 \pm 30 )$ degrees.
Most of the progress on $\gamma$ determination so far
is based on the use of the Dalitz mode, $D^0 -> K_s \pi^+ \pi^-$
\cite{ggsz}.
However, for now, this method has a disadvantage as it entails a
a modelling of the resonances involved; though model independent
methods of analysis, at least in 
principle, exist\cite{ans_b0,ggsz,ans_path}. 
The simpler modes ({\it e.g.} $K^+ \pi^-$) require more statistics
but they would not involve such modelling error as in 
the Dalitz method. Also the higher CP asymmetries in those
modes should have greater resolving power for determination of $\gamma$.
The table
shows the statistical, systematic and the resonance-model dependent
errors on $\gamma$ separately. Note that for now i do not think the model
dependent error
(around 10 degrees) ought to be added in quadrature. That is why
the combined error of $\pm 30$ degrees is somewhat inflated to reflect
that. The important point to note is that as more B's are
accumulated, more and more decay modes can be included in determination
of $\gamma$; thus for the next several years the accuracy on $\gamma$
is expected to improve faster than $1/\sqrt(N_B)$, $N_B$ being the
number of B's.

\subsection{Direct searches: Two important illustrations}

B-decays offers a wide variety of methods for searching for NP
or for BSM-CP-odd phase(s). First we will elaborate a bit on 
the following two
methods.  

\begin{itemize}

\item{} Penguin dominated hadronic final states in $b \to s$
transitions.

\item{} Radiative B-decays.

Then we will provide a brief summary of the multitude
of possibilities
that a SBF offers, in particular, for numerous important approximate
null tests (ANTs).

\end{itemize}

\subsection{Penguin dominated hadronic final states in $b \to s$
transitions}

For the past couple of years, experiments at the two B-factories
have been showing some indications of a tantalizing possibility
{\it i.e.} a BSM-CP-odd phase in penguin dominated
$b \to s$ transitions. Let us briefly recapitulate the basic idea.

\begin{figure}[t]
\vspace{9.0cm}
\includegraphics{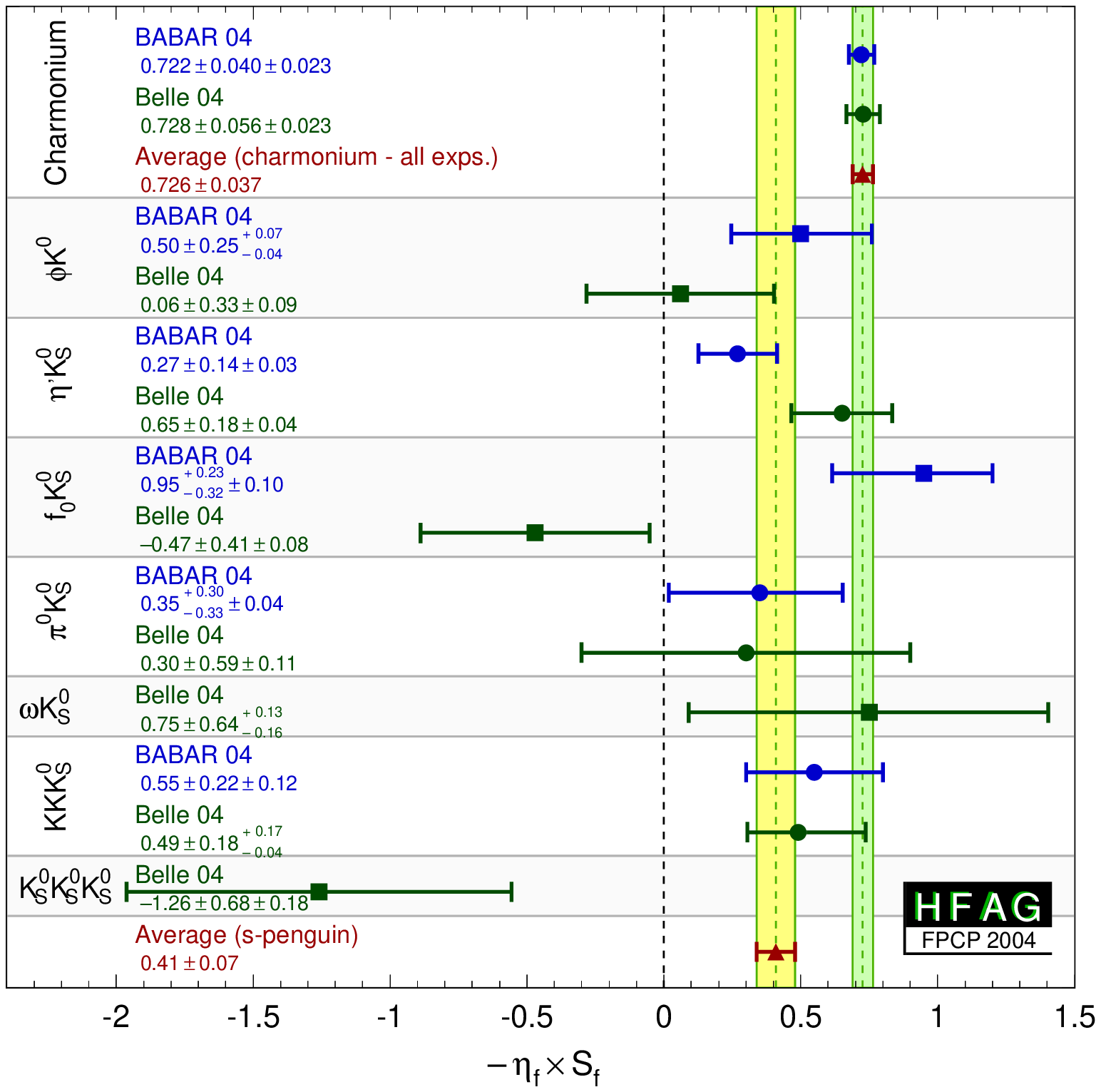}
 \caption{\it
Experimental status of $\sin 2 \beta$ from 
penguin-dominated modes; taken from ~\cite{hfag04}. 
\label{hfag_peng} }
\end{figure}

Fig.~\ref{hfag_peng} show the experimental status~\cite{hfag04}. 
With about $250 \times 10^6$ B-pairs in each of the B-factories,
there are two related
possible indications. In particular, BABAR
finds about a 3$\sigma$ deviation in $B \to \eta^{'} K_s$.
Averaging over the two experiments, this is reduced to about
2.3$\sigma$. 
Secondly adding all such
penguin dominated modes seems to indicate a 3.5$\sigma$ effect.

Since $B \to \eta^{'} K_s$ seems to be so prominently responsible
for the indications of deviations in the current data sample,
let us briefly discuss this particular FS. That the mixing induced
CP in $\eta^{'} K_s$ can be used to test the SM was 1st proposed
in ~\cite{ls97}. This was triggered in large part by
the discovery of the unexpectedly large Br for $B \to \eta^{'} K_s$.
Indeed ref. ~\cite{ls97} emphasized that the large
Br may be very useful in determining $\sin 2 \beta$
with $B \to \eta^{'} K_s$ and comparing it with the value
obtained from $B \to \psi K_s$. In fact it is precisely the
large Br of $B \to \eta^{'} K_s$ that is making the error
of the TDCP measurement the smallest amongst all the penguin dominated modes
presently studied. Note also that there is a corresponding
proposal to use the large Br of the inclusive $\eta^{'} X_s$ for
searching for NP with the use of direct CP~\cite{ans_direct,ht_97}.

Ref. ~\cite{ls97} actually suggested use of TDCP studies not just in
$\eta^{'} K_s$
but in fact also $[\eta, \pi^0, \omega, \rho, \phi...]K_s$
to test the SM. These are, indeed most of the modes
currently being used by BABAR \& BELLE.

Simple analysis in ~\cite{ls97} suggested that in all such
penguin dominated ($b \to s$) modes Tree/Penguin is small,
$< 0.04$. In view of the theoretical difficulties in reliably
estimating these effects,
Ref~\cite{ls97} emhasized that
it would be very difficult in the
SM to accomodate $\Delta S >0.10$, as a catious bound.

\subsubsection{Final state interaction effects}

The original papers~\cite{gw97,rf97,ls97} predicting,

\be
\Delta S_f = S_f - S_{\psi K} \approx 0
\en
used naive factorization; in particular, FSI were completly
ignored. A remarkable discovery of the past year is that
in several charmless 2-body B-decays direct CP asymmetry is rather large.
This means that FSI (CP-conserving) phase(s)
in exclusive B-decays
need not be small~\cite{ccs1}. Since these are 
non-perturbative~\cite{bss_pert}, model dependence
becomes unavoidable. Indeed characteristically these FSI
phase(s) arise formally from $O(1/m_B)$ corrections:

\begin{itemize}

\item{} In pQCD~\cite{pqcd} a phenomenological parameter $k_T$,
corresponding to the transverse momentum of partons, is introduced
in order to regulate the end point
divergences encountered in power corrections. 
This in turn gives rise to sizable strong phase difference from
penguin induced annihilation.

\item{} In QCDF~\cite{qcdf}, in its nominal version, the direct CP
asymmetry in many channels (e.g $B^0 \to K^+ \pi^-, \rho^- \pi^+,
\pi^+ \pi^-$.....) has the opposite sign compared to the
experimental findings. Just like in the pQCD
approach where the annihilation topology play an important role
in giving rise to large strong phases, and for explaining the 
penguin-dominated VP modes, it has been suggested in~\cite{qcdf_s4}
that in a specific scenario (S4), for QCDF to
agree with the Br 
of penguin-dominated PV modes as well as with the measured sign of
the direct asymmetry in the prominent channel 
$B^0 \to K^+ \pi^-$, a large annihilation
contribution be allowed by choosing $\rho_A =1$, $\phi_A= -55^\circ$
for PP, $\phi_A = -20^\circ$ for PV and $\phi_A = -70^\circ$
for VP modes. 

\item{} In our approach~\cite{ccs1}, 
QCDF is used for short-distance (SD) physics;
however, to avoid double-counting, we set the above two parameters
[$\rho_A$, $\phi_A$] as well as two additional parameters
[$\rho_H$, $\phi_H$] that they have~\cite{qcdf_s4}
to zero. Instead we try to include long-distance ($1/m_B$) corrections
by using on-shell rescattering of 2-body modes 
to give rise to the needed FS phases.

\end{itemize}

So, for example,
color-suppressed modes such as $B^0 \to K^0 \pi^0$ gets
important contributions from color allowed processes:
$B^0 \to K^{-(*)} \pi^{+}(\rho^+), D_S^{-(*)} D^{+(*)}$.
The coupling strengths at the three vertices of such a triangular
graph are chosen to give the known rates of corresponding
physical processes such as $B^0 \to D_S^{-*} D^{+(*)}$,
$D^* \to D + \pi$ etc. Furthermore, since these vertices are not
elementary and the exchanged particles
are off-shell, form-factors have to be introduced so that loop
integrals become convergent.
Of course, there is no way to determine
these reliably.
We vary these as well as other parameters
so that Br's are in rough agreement with experiment, 
then we calculate the
CP-asymmetries.  

Recall the standard form for the asymmetries:\\
\be
 {\Gamma(\ov B(t)\to f)-\Gamma(B(t)\to f)\over
  \Gamma(\ov B(t)\to f)+\Gamma(B(t)\to
   f)}={\cal S}_f\sin(\Delta mt)+\A_f\cos(\Delta mt)
\en

The TDCP asymmtery ($S_f$) and direct CP 
asymmetry [$A_f \equiv -C_f$ (BaBar notation)] both depend on
the strong phase. Thus measurements of direct CP asymmetry $A_f$ (in
addition to $S_f$) allows tests of model calculations, though in
practice its real use may be limited to those cases where the
direct CP asymmetry is not small. This is the case, for example,
for $\rho^0 \ks$ and  $\omega \ks$~\cite{ccs2}.

It is also important to realize that not only there is a
correlation between $S_f$ and $A_f$ for FS in $B^0$ decays, but also that
the model entails specific predictions for direct CP in the charged
counterparts. So, for example, in our model for FSI,
large direct CP asymmetry is also expected in the charged counterparts
of the above two modes.

In addition to two body modes there are also very interesting 3-body
modes such as $B^0 \to K^+ K^- K_S(K_L), K_S K_S K_S (K_L)$.
These may also be useful to search for NP as they are also penguin
dominated. We use resonance-dominance of the relevant two body
channels to extend our calculation of LD rescattering phases in
these decays~\cite{ccs3}.

Tables ~\ref{mix1_tab} and ~\ref{mix2_tab} summarize our results for
$\Delta S$ and A for two body and 3-body modes.
We find that~\cite{ccs2,ccs3} $B^0 \to \eta^{'} K_S$, $\phi K_S$ and 3$K_S$ are
cleanest~\cite{mb05}, i.e. central values of $\Delta S$ as well as the errors
are rather small, O(a few\%). Indeed we find that even after including
the effect of
FSI, $\Delta S$ in most of these penguin-dominated modes,  
it is very difficult to get $\Delta S >0.10$ in the SM. Thus we can
reiterate (as in~\cite{ls97}) that $\Delta S >0.10$ would be a strong
evidence for NP.

\begin{table}[t]
\caption{Direct CP asymmetry parameter $\A_f$ and the
mixing-induced CP parameter $\Delta S_f^{SD+LD}$ for various
modes. The first and second theoretical errors correspond to the
SD and LD ones, respectively (see~\cite{ccs2} for details). The
$f_0K_S$ channel is not included as we cannot make reliable
estimate of FSI effects on this decay; table adopted 
from~\cite{ccs2}.} \label{mix1_tab}
\begin{tabular}{l r c r c} \hline  
&  \multicolumn{2}{c}{$\Delta S_f$}
&   \multicolumn{2}{c}{$\A_f(\%)$}  \\ \cline{2-3} \cline{4-5}
\raisebox{2.0ex}[0cm][0cm]{Final State} & SD+LD & Expt &
SD+LD & Expt
\\ \hline
$\phi K_S$ & 
$0.03^{+0.01+0.01}_{-0.04-0.01}$ & $-0.38\pm0.20$  &
$-2.6^{+0.8+0.0}_{-1.0-0.4}$  & $4\pm17$ \\
$\omega K_S$ &
$0.01^{+0.02+0.02}_{-0.04-0.01}$  & $-0.17^{+0.30}_{-0.32}$
& $-13.2^{+3.9+1.4}_{-2.8-1.4}$ &
$48\pm25$ \\
$\rho^0K_S$ &
$0.04^{+0.09+0.08}_{-0.10-0.11}$ & -- &
$46.6^{+12.9+3.9}_{-13.7-2.6}$ & -- \\
$\eta' K_S$ &
$0.00^{+0.00+0.00}_{-0.04-0.00}$
& $-0.30\pm0.11$  &
$2.1^{+0.5+0.1}_{-0.2-0.1}$ & $4\pm8$ \\
$\eta K_S$ &
$0.07^{+0.02+0.00}_{-0.05-0.00}$ & $-$
& $-3.7^{+4.4+1.4}_{-1.8-2.4}$ & $-$
\\
$\pi^0K_S$ & 
$0.04^{+0.02+0.01}_{-0.03-0.01}$ & $-0.39^{+0.27}_{-0.29}$
&
$3.7^{+3.1+1.0}_{-1.7-0.4}$ &
$-8\pm14$  \\
\hline
\end{tabular}
\end{table}


\begin{table}[t]
\centering
\caption{Mixing-induced and direct CP asymmetries $\sin
2\beta_{\rm eff}$ (top) and $\A_f$ (bottom), respectively, in
$B^0\to K^+K^-K_S$ and $K_SK_SK_S$ decays. Results for
$(K^+K^-K_L)_{CP\pm}$ are identical to those for
$(K^+K^-K_S)_{CP\mp}$; table taken from ~\cite{ccs3}}. 
\label{mix2_tab}
\begin{tabular}{l r r} \hline  
Final State & $\sin 2\beta_{\rm eff}$  & Expt.  \\
\hline
$(K^+K^-K_S)_{\phi K_S~{\rm excluded}}$
             & $0.749^{+0.080+0.024+0.004}_{-0.013-0.011-0.015}$
	                 & $0.57^{+0.18}_{-0.17}$
			             \\
$(K^+K^-K_S)_{CP+}$
				                  &
						  $0.770^{+0.113+0.040+0.002}_{-0.031-0.023-0.013}$
						              &
	                  \\
$(K^+K^-K_L)_{\phi
	   K_L~{\rm
									   excluded}}$
									               &
										       $0.749^{+0.080+0.024+0.004}_{-0.013-0.011-0.015}$
										                   &
												   $0.09\pm0.34$
												               \\
													        $K_SK_SK_S$
														            &
															    $0.748^{+0.000+0.000+0.007}_{-0.000-0.000-0.018}$
															                &
																	$0.65\pm0.25$
																	            \\
																		     $K_SK_SK_L$
																		                 &
																				 $0.748^{+0.001+0.000+0.007}_{-0.001-0.000-0.018}$
																				             &
																					                 \\
																							  \hline
																							    &$\A_f(\%)$
																							    &Expt.
																							    \\
																							     \hline
																							      $(K^+K^-K_S)_{\phi
																							      K_S~{\rm
																							      excluded}}$
																							                  &
																									  $0.16^{+0.95+0.29+0.01}_{-0.11-0.32-0.02}$
																									              &
																										      $-8\pm10$
																										                  \\
																												   $(K^+K^-K_S)_{CP+}$
																												               &
																													       $-0.09^{+0.73+0.16+0.01}_{-0.00-0.27-0.01}$
																													                   &
																															               \\
																																        $(K^+K^-K_L)_{\phi
																																	K_L~{\rm
																																	excluded}}$
																																	            &
																																		    $0.16^{+0.95+0.29+0.01}_{-0.11-0.32-0.02}$
																																		                &
																																				$-54\pm24$
																																				            \\
																																					     $K_SK_SK_S$
																																					                 &
																																							 $0.74^{+0.02+0.00+0.05}_{-0.06-0.01-0.06}$
																																							             &
																																								     $31\pm17$
																																								                 \\
																																										  $K_SK_SK_L$
																																										              &
																																											      $0.77^{+0.12+0.08+0.06}_{-0.28-0.11-0.07}$
																																											                  &
																																													              \\

%
\hline
\end{tabular}
\end{table}

Having said that, it is still important to stress that genuine
NP in these penguin dominated modes must show up in many
other channels as well. Indeed, on completely model independent
grounds~\cite{brwns}, the underlying NP has to be 
either in the 4-fermi vertex
(bss$\bar s$) or (bsg, $g=gluon$).  In either case, it has to
materialize into a host of other reactions and phenomena and it is not
possible that it only effects time dependent CP in say $B \to \eta' K_s$
and/or $\phi K_s$ and/or 3$K_s$. For example, for the 4-fermi
case, we should also
expect non-standard effects in $B_d \to \phi(\eta^{'}) K^*$, $B^+ ->
\phi (\eta^{'}) K^{+(*)}$, $B_s \to \phi \phi (\eta^{'})$...In the
second case not only there should be non-standard effects in these
reactions but also in $B_{d(u)} \to X_s \gamma$, $K^* \gamma$, $B_s \to \phi
\gamma$..... and also in the corresponding $l^+ l^-$ modes. Unless
corroborative  evidence is seen in many such processes,
the case for NP due to the non-vanishing of $\Delta S$ is
unlikely to be compelling, especially if (say) $\Delta S \lsim 0.15$.

\subsubsection{Averaging issue}
As already emphasized in ~\cite{ls97}, to the extent that penguin
contributions dominate in these many modes
and $tree/penguin$ is only a few percent testing the SM by adding
$\Sigma \Delta S_f$, where $f = K_S + \eta^{'} (\phi, \pi, \omega, \rho,
\eta, K_S K_S$...), is sensible at least from a theoretical standpoint.
At the same time it is important to emphasize that a convincing case
for NP requires unambiguous demonstration of significant
effects (i.e. $\Delta S >0.10$) in several individual channels.

\subsubsection{Sign of $\Delta S$}

For these penguin-dominated modes, $\Delta S_f$ is primarily
proportional to the hadronic matrix element $<f | \bar u \Gamma
b \bar s \Gamma' u|B^0>$.
Therefore, in the SM for several of the final states (f),
$\Delta S_f$ could have the same sign. So a systematic trend of
$\Delta S_f$ being positive or negative ({\it and small of O(a few \%)})
does not necessarily mean NP.

The situation wrt to $\eta' K_S$ is especially interesting.
As has been known for the past many years this mode has a very large Br,
almost a factor
of 7 larger than the similar two body K $\pi$ mode.
This large Br is of course also the reason  why the statistical
error is the smallest, about a factor of two less than any other
mode being used in the test. For this reason, it is gratifying
that $\eta' K_S$ also happens to be theoretically very clean in
several of the model calculations. This has the important repercussion
that confirmation
of a significant deviation from the SM,
may well come 1st by using the $\eta' K_S$ mode, perhaps
well ahead of the other modes~\cite{js}.

\subsubsection{Concluding remarks on penguin-dominated modes}

Concluding this section we want to add that while at present there is no
clear or compelling deviation from the SM
the fact still remains that this is a very important
 approximate null test (ANT). It is exceedingly important to follow this test
 with the highest luminosity possible to firmly establish that
 as expected in the SM, $\Delta S_f$ is really \lsim $0.05$ and is not
 significantly different from this expectation. To establish this
 firmly,
 for several of the modes of interest, may well require a SBF.

\section{Time dependent CP in exclusive radiative B-decays}

Br ($B \to \gamma X_{s(d)}$) and direct CP asymmetry
$a_{cp}(B \to \gamma X_{s(d)})$ 
are well known tests of the SM~\cite{hurth_rmp, hiller_fpcp, kn98,ksw00}.
Both of these use the inclusive reaction where the theoretical prediction
for the SM are rather
clean; the corresponding exclusive cases are theoretically problematic
though experimentally more accessible.
In 1997 another important test~\cite{ags} of the SM was
proposed which used mixing induced CP (MICP) or time-dependent CP
(TDCP) in exclusive modes such as $B^0 \to K^* \gamma, \rho \gamma$.....
This is based on the simple observation that in the SM, photons produced
in reactions such as $B \to K^* \gamma, K_2^* \gamma, \rho \gamma$...
are predominantly right-handed whereas those in $\bar B^0$ decays are
predominantly left-handed. To the extent that FS of $B^0$ and $\bar B^0$
are different MICP would be suppressed in the SM.
Recall, the LO $H_{eff}$ can be written as
\be
H_{\rm eff} = - \sqrt{8} G_F \frac{e m_b}{16\pi^2} F_{\mu\nu}
\left [
    F_L^q\ \OL q\sigma^{\mu\nu}\frac{1+\gamma_5}{2}b
        +
        F_R^q\ \OL q\sigma^{\mu\nu}\frac{1-\gamma_5}{2}b
          \right ] + h.c.
            \label{effective_H}
\en
            \noindent Here $F_L^q$ ($F_R^q$) corresponds to the
            amplitude for the
            emission of left (right) handed photons in the $b_R
            \to q_L \gamma_L$
            ($b_L \to q_R \gamma_R$) decay, {\it i.e.} in the
            $\OL B \to \OL F
            \gamma_L$ ($\OL B \to \OL F \gamma_R$) decay.
\subsection{Application to $\bo,\bs \to$ vector meson + photon}
Thus, based on the SM, {\it LO $H_{eff}$},  
in b-quark decay (i.e. $\bar B$ decays), the amplitude for
producing wrong helicity (RH) photons $\propto m_q/m_b$ where $m_q
=m_s$ or $m_d$ for $b \to s \gamma$ or $b \to d \gamma$
respectively. Consequently the TDCP asymmetry is given by,

\be
\bo \to K^{*0}\gamma &:& A(t) \approx (2m_s/m_b)\sin(2\beta)\sin(\Delta
mt)~,
\nonumber\\
\bo \to \rho^0\gamma &:& A(t) \approx 0~,
\nonumber\\
\bs \to \phi\gamma &:& A(t) \approx 0~,
\nonumber\\
\bs \to K^{*0}\gamma &:& A(t) \approx -(2m_d/m_b)\sin(2\beta)\sin(\Delta
mt)~,
\label{examples}
\en
where $K^{*0}$ is observed through $K^{*0}\to K_S \pi^0$.

Interestingly not only emission of wrong-helicity photons from
B decays is highly suppressed, in many extensions of the SM,
{\it e.g.} Left-Right Symmetric models (LRSM) or 
SUSY~\cite{chn03,
okada03,chl00}
or Randall-Sundrum (warped extra dimension~\cite{aps04}) models, 
in fact they can be enhanced by the ratio $m_{heavy}/m_b$ where
$m_{heavy}$ is the mass of the virtual fermion in the penguin-loop.
In LRSM as well as some other extensions this enhancement can
be around $m_t/m_b$. So while in the SM the asymmetries
are expected to be very small, they can be sizeable in LRSM~\cite{ags}
(see Table~\ref{ags_tab}) as well as in many other models.

\begin{table}
\begin{center}
\hspace*{-.2in}
\begin{tabular}{|c|c|c|}
\hline
Process & SM & LRSM \\
\hline
$A(B \to K^* \gamma)$ & $2\frac{m_s}{m_b}
\sin 2\beta \sin(\Delta m_t)$ &
$\sin 2 \omega \cos 2 \beta \sin (\Delta m_t)$ \\
\hline
$A(B \to \rho \gamma)$ & $\approx 0$ & $\sin 2 \omega \sin (\Delta m_t)$
\\
\hline
\end{tabular}
\end{center}
\caption{Mixing-induced CP asymmetries in radiative exclusive
B-decays in the SM and in the LRSM. Note $|\sin 2 \omega| \lsim 0.67$
is allowed~\cite{ags,brwns}}
\label{ags_tab}
\end{table}

\subsection{Generalization to $\bo, \bs \to$ two pseudoscalars + photon}

An important generalization was made in Ref~\cite{aghs}. It was
shown that the basic validity of this test of the SM does not require
the final state to consist of a spin one meson (a resonance
such as $K^*$ or $\rho$) in addition to a photon.
In fact the hadronic final states can equally well be two 
mesons; {\it e.g.} $\ks (\pi^0, \eta', \eta, \phi...)$ or
$\pi^+ \pi^-$.
Inclusion of these non-resonant final states, in addition to the
resonances clearly enhances the sensitivity of the test considerably.
For the case when the two mesons are antiparticle of each other
{\it e.g. }
$\pi^+ \pi^-$, then there is the additional advantage that both the magnitude
and the weak phase of any new physics contribution may be determined
from a study of the angular distribution~\cite{aghs}.

\subsection{Theoretical subtelties}

In principle, photon emission from the initial light-quark
is a non-perturbative, long-distance, contamination
to the interesting signal of the short-distance
dipole emission from $H_{eff}$~\cite{abs,grinpir}. Fortunately, it
can be shown~\cite{aghs} that predominantly these LD
photons have the same helicity as those from $H_{eff}$.

Another important source of SM contamination was recently emphasized in
Ref.~\cite{gglp} from processes such as $b \to s \gamma + $ gluon
which are from non-dipole operators. Such processes do not fix the
helicity of the photon and so can make a non-vanishing SM
contribution to mixing induced CP.

It was emphasized in Ref~\cite{aghs} that the presence of
such non-dipole contributions can be separated from
the dipole contributions, though, it may require
larger amount of data, the resolution
to this problem is
data driven.

To briefly recapitulate, 
the different operator structure in $H_{eff}$
would mean, that in contrast to the pure dipole case,
the time dependent CP asymmetry (S) would be a function of the
Dalitz variables, the invariant mass (s) of the meson pair,
and the photon angle of emission (z). 
A difference in the values
of S for two resonances of identical $J^{PC}$ would also
mean presence of non-dipole contributions. Schematically,
we may write:

\be
d S^i/(ds dz) & = & [A_{\sigma} + A_0^i] + B^i s + C^i z
\en
where $A_{\sigma}$ is the ``universal" contribution that one
gets from the dipole operator of the $H_{eff}$ no matter if it is 
a resonance, or a non-resonance mode. It is distinct from
the contribution of the 4-quark operators as not only
it is independent of energy (s) or angle (z) Dalitz variables
but also it is independent of the specific nature of the hadronic
FS ({\it i.e.} resonant or non-resonant). The remaining
contributions are all originating from 4-quark operators;
not only they dependent on energy and angle but also the 
coefficients are expected to vary from one FS to
another. In particular the 4-quark operators may give a FS dependent
(energy and angle independent) constant $A_0^i$.  
It is easy to convince oneself that with sufficient data
the important term $A_{\sigma}$, at least in principle, can be separated.
Once that is done its size should be indicative of whether it is
consistent with expectations from SM or requires new physics 
to account for it.  

\subsection{Approximate null tests aglore!}

If the effects of a BSM CP-odd phase on B-physics are small, then
searching for these via {\it \bf null tests} becomes especially
important. Since CP is not an exact symmetry of the SM, it is very
difficult if not impossible to find exact null tests. Fortunately
clean environment at a SBF should allow many interesting
approximate null tests (ANTs); see Table~\ref{ants_tab}~\cite{brwns}.

Clearly there is a plethora of powerful tests for a new
CP-odd phase and /or new physics that a SBF should allow us
to do. Perhaps especially noteworthy (in addition
to penguin-dominated hadronic and radiative
B decays) are the numerous very
interesting tests pertaining to $B \to X(K,K^*..) l^+ l^-$ 
~\cite{hurth_rmp,hiller_fpcp}. 

Furthermore search for the transverse polarization~\cite{aes_tau,gl_tau}
of the
$\tau$ in $B \to X(D,D^*..) \tau \nu_{\tau}$ due
to their unique cleanliness are extremely interesting especially
in light of the discovery of neutrino mass and the potential
richness of neutrinos with the possible
presence of Majorana neutrinos in simple grounds-up
extensions of the SM as well as in many 
other approaches~\cite{abs_3g,rnm}.

Sensitivity of each of these to NP as well as theoretical
cleanliness ({\it i.e.} how reliable SM predictions are) for each
is also indicated. It should be clear that for most of these tests
$> 5 \times 10^9$ B-pairs are essential, that is a SBF.


\begin{table}[t]  
\caption{Final states and observables in B - decays
useful in searching for effects of New Physics. Reliability of
SM predictions ({\it i.e.} how clean) and sensitivity to new physics
are each indicated by
stars ($5 = best$); table adopted from~\cite{brwns}}.
\begin{tabular}{|c|c|c|c|}
\hline
Final State&Observable&how clean&how sensitive\\  
\hline
$\gamma [K_s^*,\rho,\omega]$ & TDCP & 5* & 5*  \\
\hline
$K_s[\phi,\pi^0,\omega,\eta',\eta,\rho^0]$ & TDCP & 4.5* &  5* \\
$K^*[\phi,\rho,\omega]$ & TCA &  4.5* &  5* \\
\hline
$[\gamma,l^+l^-] [X_s,X_d]$ & DIRCP & 4.5* & 5*  \\
\hline
same   & Rates & 3.5* & 5* \\
\hline
$\jpsi K $ & TDCP, DIRCP & 4* & 4*  \\
\hline
$\jpsi K^* $ & TCA &  5* & 4*  \\
\hline
$D (*) \tau \nu_{\tau}$ & TCA ($p_t^{\tau})$ & 5* & 4*  \\
same & Rate & 4* & 4* \\
\hline
\hline
\end{tabular}
\label{ants_tab}
\end{table}


\section{K-Unitarity Triangle}

For the past many years, effort has been directed towards
constraining the UT especially the parameters $\rho$ and $\eta$
by a combination of information from K and B-physics, as
mentioned briefly in Section 1. With the advent of B-factories
and significant advance that has been already made (and
a lot more is expected to come) it has become possible
to construct the UT purely from B-physics~\cite{ckmfit,utfit}.
In fact it may also be very interesting and important to
construct a separate UT from K-decays. This could become
particularly useful in  search for small deviations.
Reactions that are relevant for a K-UT are~\cite{pascos03}:

\begin{itemize}

\item{} Indirect CP-violation parameter, $\epsilon_K$ with
the hadronic matrix elements (parameter $B_K$) from the lattice.
With the dawning of the era of dynamical
simulations using discretizations that
preserve chiral-flavor symmetries of the continuum~\cite{rbc_nf2},
lattice should be able to significantly reduce the errors
on $B_K$~\cite{rbc_jun}.

\item{} Accurate measurements of the BR
of $K ^+ \to \pi^+ \nu \bar \nu$ can give a clean
determination of $ |V_{td}|$~ \cite{ajb_nunu}. 
Important progress has been recently made in the 1st
step towards an accurate determination of this Br~\cite{e787_bnl}.  
Charm quark contribution in the penguin
graph is difficult to reliably estimate but this is expected
to be subdominant~\cite{flp_nunu}.

\item{}
 Measurement of the BR of $K_L \to \pi^0 \nu \bar \nu$
can give an extremely clean value of $\eta$, {\it i.e.} $Im
V_{td}$. This is clearly very challenging experimentally; however,
it is unique in its cleanliness, perhaps on the same footing as
$\gamma$ from $BKD$ processes discussed above.  

\item{}
After enormous effort, the experimentalists have determined
the direct CP violation parameter $\epsilon'/\epsilon$ 
with considerable accuraccy~\cite{na48,ktev}.
For theory a reliable calculation remains a very important 
outstanding challenge.
Recently it has become clear that not only chiral symmetry on
the lattice is essential for this calculation but also the quenched
approximation suffers here from very serious
pathology~\cite{gp,jlq6}. 
As mentioned above, since the past 2-3 years 
considerable effort is being expended in generation of 
dynamical configurations with domain wall quarks
which possess excellent chiral properties. In the near future
we should expect to see the application of these
new generation of lattices for study of $\epsilon'/\epsilon$.
It remains to be seen as to
how accurately the current generation of computers can allow
this important calculation to be done.

\end{itemize}

\section{Neutron electric dipole moment: a classic ANT of the SM}

In the SM, neutron electric dipole moment (nedm)
cannot arise at least to two EW loops; thus is expected to be
exceedingly small, {\it i.e.} $\lsim 10^{-31}ecm$. Long series of
experiments over the past several decades now place a 90 \%CL
bound of $\lsim 6.3 \times 10^{-26}$ecm~\cite{nedm_bound}. 
So the expectation
from the SM is many orders of magnitude below the current
experimental bound. In numerous extensions of the SM,
including SUSY, warped extra dimensions etc. nedm close
to or even somewhat bigger than the current experimental bound
occurs~\cite{bgk_nedm, aps04}. 
Thus continual experimental improvements of this bound remains
a very promising way to discover new BSM CP-odd phase(s).

\section{Top quark electric dipole moment: another
clean null test of the SM}

The top quark is so heavy compared to the other quarks that the
GIM-mechanism is extremely effective. Thus in the decays of the
top-quark, in the SM, all FCNC are extremely suppressed. Once
again, top quark edm cannot arise in the SM to two EW loops
and  is therefore expected to be extremly small. In many BSM
scenarios with extra Higgs doublets~\cite{xs,bsp}, 
LRSM, SUSY~\cite{abes_pr}, the top quark can
acquire edm at one loop and consequently can be 
considerably bigger (See Table~\ref{dipolesumtable}). Therefore searches for  
the top dipole moment at the International Linear Collider will be
an important goal~\cite{abes_pr,lc_book}. Indeed if sufficient high luminosity could be
attained top quark edm of around $10^{-19}$ ecm may well be
detectable (See Table~\ref{tdm_at_ilc}).   

\begin{table}
\begin{center}
\caption{Expectations for top edm form-factor in SM and beyond;
adopted from~\cite{abes_pr}}
\bigskip
\protect\label{dipolesumtable}
\begin{tabular}{|r||r||r|r|r|r|} \cline{2-5}
\hline \hline
type of moment & $\sqrt s$ & Standard&Neutral Higgs 
& Supersymmetry \\
$(e-cm)~~ \Downarrow$ &$({\rm GeV}) \Downarrow$ & Model
& $m_{h}=100 -300$ &$m_{\tilde g}=200-500$ \\
\hline
\hline
&500& & $(4.1-2.0)\times 10^{-19}$ 
& $(3.3-0.9)\times 10^{-19}$ \\
$|\Im{\rm m}(d_t^\gamma)|$& & $< 10^{-30}$ & & \\
& 1000& & $(0.9-0.8)\times 10^{-19}$ 
& $(1.2-0.8)\times 10^{-19}$ \\ \hline
\hline
&500& & $(0.3-0.8)\times 10^{-19}$
& $(0.3-0.9)\times 10^{-19}$ \\
$|\Re{\rm e}(d_t^\gamma)|$& & $< 10^{-30}$ & & \\
& 1000 & & $(0.7-0.2)\times 10^{-19}$
& $(1.1-0.3)\times 10^{-19}$ \\ \hline
\hline
&500& &$(1.1-0.2)\times 10^{-19}$ 
& $(1.1-0.3)\times 10^{-19}$  \\
$|\Im{\rm m}(d_t^Z)|$& & $< 10^{-30}$ & & \\
& 1000& & $(0.2-0.2)\times 10^{-19}$
&$(0.4-0.3)\times 10^{-19}$   \\ \hline
\hline
&500& & $(1.6-0.2)\times 10^{-19}$ 
& $(0.1-0.3)\times 10^{-19}$ \\
$|\Re{\rm e}(d_t^Z)|$& & $< 10^{-30}$ & & \\
& 1000& &$(0.2-1.4)\times 10^{-19}$
& $(0.4-0.1)\times 10^{-19}$ \\ \hline
\hline
\end{tabular}
\end{center}
\end{table}


\begin{table}
\begin{center}
\caption[first entry]{\emph{Attainable 1-$\sigma$ sensitivities
to the CP-violating
dipole moment form factors in units of $10^{-18}$ e-cm, with ($P_e=\pm
1$)
and without ($P_e=0$)
beam polarization. $m_t=180$ GeV\null. Table
taken from \cite{abes_pr,bbo_tdm}. \protect\label{tdm_at_ilc}}}
\bigskip

\begin{tabular}{|c|c|c|c||c|c|c|}
\hline
  & \multicolumn{3}{c||}{$20 {\mbox{\ fb}}^{-1},\sqrt{s}=500{\mbox{\
  GeV}}$} &
  \multicolumn{3}{c|} {$50 {\mbox{\ fb}}^{-1},\sqrt{s}=800{\mbox{\
  GeV}}$} \\
    & $P_e=0$ & $P_e=+1$ & $P_e=-1$ & $P_e=0$ & $P_e=+1$ &  $P_e=-1$ \\
    \hline
    $\delta(\Re{\rm e}d_{t}^{\gamma})$ & 4.6 & 0.86 & 0.55 & 1.7 & 0.35
    & 0.23 \\
    \hline
    $\delta(\Re{\rm e}d_{t}^{Z})$ & 1.6 & 1.6 & 1.0  & 0.91 & 0.85 &
    0.55 \\
    \hline
    $\delta(\Im{\rm m}d_{t}^{\gamma})$ & 1.3 & 1.0 & 0.65 & 0.57 & 0.49
    & 0.32 \\
    \hline
    $\delta(\Im{\rm m}d_{t}^{Z})$ & 7.3 & 2.0 & 1.3  & 4.0 & 0.89 & 0.58
    \\
    \hline
    \end{tabular}
    \end{center}
    \end{table}

\section{Summary}

The new millennium marks the spectacular success of B-factories
leading to a milestone in our understanding of CP-violation;
in particular, for the first time CKM paradigm
of CP violation is quantitatively confirmed.

Direct measurement of $\sin2\beta$ by the B-factories agrees
remarkably well with the theoretical expectation from the SM to
about ~10\%.  Furthermore, first relatively crude {\it direct}
determination of the other two angles ($\alpha$ \& $\gamma$) also
are consistent with theoretical expectations. While these findings
are good news for the SM, at the same ime, they imply that most
likely the effect of BSM CP-odd phase on B-physics is likely to be a
small perturbation. Thus discovery of new BSM-CP-odd source(s) of
CP violation in B-physics is likely to require very large, clean,
data samples and extremely clean predictions from theory.

For the search of such small deviations approximate 
null tests of the SM gain
new prominence.

Also important for this purpose is the drive to directly determine
all three angles of the UT with highest precision possible, {\it
i.e.} with errors roughly around the errors
allowed by theory. It should be clear that to accomplish
this important goal would require a Super-B Factory.

Specifically regarding penguin-dominated hadronic FS,
that have been much in the recent news, the current data does not show
any convincing signal for deviation from the SM; however, it is
a very important and sensitive test for new physics 
and its of vital importance
to
reduce the experimental errors to O(5\%); for this purpose too a SBF may well
be needed.

Outside of B-Physics, K-unitarity triangle,
neutron electric dipole moment and top quark dipole moment
are also very important {\it approximate null tests} of the SM that 
should be pursued
vigorously.

\section{Acknowledgements}
I want to thank the organizers (and, in particular, Mario Greco)
for their kind invitation. Useful discussions with
David Atwood, Tom Browder, Chun-Khiang Chua,
Tim Gershon, Masashi Hazumi, Jim Smith 
and especially Hai-Yang Cheng 
are gratefully acknowledged. This research was supported in part by
DOE contract No. DE-FG02-04ER41291.
\end{document}